# Nitrogen Decoration of Basal Plane Dislocations in 4H-SiC

Jiajun Li[1,2], Hao Luo[1,2], Guang Yang[3], Yiqiang Zhang[4],

Xiaodong Pi[1,2*], Deren Yang[1,2], Rong Wang[1,2*]

[1]*State Key Laboratory of Silicon Materials and School of Materials Science and Engineering, Zhejiang University, Hangzhou, 310027, China*

[2]*Institute of Advanced Semiconductors & Zhejiang Provincial Key Laboratory of Power Semiconductor Materials and Devices, Hangzhou Innovation Center, Zhejiang University, Hangzhou, 311200, China*

[3]*Key Laboratory of Optical Field Manipulation of Zhejiang Province, Department of Physics, Zhejiang Sci-Tech University, Hangzhou 310018, China*

[4]*School of Materials Science and Engineering & Henan Institute of Advanced Technology, Zhengzhou University, Zhengzhou, Henan 450001, China*

**Abstract**

Basal-plane dislocations (BPDs) pose a great challenge to the reliability of bipolar power devices based on the 4H silicon carbide (4H-SiC). It is well established that heavy nitrogen (N) doping promotes the conversion of BPDs to threading edge dislocations (TEDs) and improves the reliability of 4H-SiC-based bipolar power devices. However, the interaction between N and BPDs, and the effect of N on the electronic properties of BPDs are still ambiguous, which significantly hinder the understanding on the electron-transport mechanism of 4H-SiC-based bipolar power devices. Combining molten-alkali etching and the Kelvin probe force microscopy (KPFM) analysis, we demonstrate that BPDs create acceptor-like states in undoped 4H-SiC, while acting as donors in N-doped 4H-SiC. First-principles calculations verify that BPDs create occupied defect states above the valence band maximum (VBM) and unoccupied defect states under the conduction-band minimum (CBM) of undoped 4H-SiC. The electron transfer from the defect states of intrinsic defects and native impurities to the unoccupied defect states of BPDs gives rise to the acceptor-like behavior of BPDs in undoped 4H-SiC. Defect formation energies indicate that N atoms can spontaneously decorate BPDs during the N doping of 4H-SiC. The binding between N and BPD is strong against decomposition. The accumulation of N dopants at the core of BPDs results in the accumulation of donor-like states at the core of BPDs in N-doped 4H-SiC. This work not only enriches the understanding on the electronic behavior of BPDs in N-doped 4H-SiC, but also helps understand the electron transport mechanism of 4H-SiC-based bipolar power devices.

*Corresponding authors.

E-mail address: xdpi@zju.edu.cn; rong_wang@zju.edu.cn

## I. INTRODUCTION

4H silicon carbide (4H-SiC) has been attracting great attention in high-power and high-frequency electronics, owing to the advantages of wide bandgap, high breakdown electric field strength, and high thermal conductivity [1,2]. The explosive development of electric vehicles, 5G communications, and low-loss power transmission systems has inspired intense researches on high-quality 4H-SiC, one of the most significant challenges is to reduce the density of dislocations in 4H-SiC [3,4]. With the hexagonal structure of 4H-SiC, dislocations in 4H-SiC can be classified into threading dislocations (TDs) and basal-plane dislocations (BPDs) [5,6]. The Burgers vector of a perfect BPD in 4H-SiC is $[11\bar{2}0]a/3$. The BPD is prone to dissociate into partial dislocations (PDs) by $[11\bar{2}0]a/3 \rightarrow [10\bar{1}0]a/3 + [01\bar{1}0]a/3$ [7,8]. The decomposed BPD consists of two PDs with the Si core and C core, which are separated by a single Shockley-type stacking fault (SSF) on the basal plane of 4H-SiC [9]. During the operation of 4H-SiC-based bipolar devices, one of the most severe reliability issues is the forward-voltage degradation that has been attributed to the expansion of the SSF activated by the UV illumination or the electron-hole recombination [10-12]. Therefore, reducing the density of BPDs is extremely useful to improve the reliability of 4H-SiC-based electronic devices.

The generation mechanism and evolution of BPDs have been investigated by experimental and theoretical researches. It has been found that the thermal stress during the growth of 4H-SiC single crystals introduces high density of BPDs in 4H-SiC [13-15]. Furthermore, processing of 4H-SiC wafers, including slicing, lapping, and chemical mechanical polishing (CMP), also create BPDs in 4H-SiC as a result of processing-induced shear stress [16–19]. Molten-alkali etching is usually adopted to reveal BPDs and facilitate the statistics of BPD density in off-axis sliced n-type 4H-SiC substrates [20-22]. After years of development, the density of BPDs in 4H-SiC substrate wafers is reduced to the order of magnitude of $10^2$-$10^3$ cm$^{-2}$ [23]. The high-density BPDs can be converted to threading-edge dislocations (TEDs) during the homoepitaxy of 4H-SiC, which can be promoted by various approaches such as increasing the growth rate, high-temperature annealing, heavily nitrogen (N) doping, lateral growth via etch pits of BPDs, and growth interruptions [6,24-27]. Among these approaches, inserting a heavily N-doped buffer layer has been widely adopted to reduce the density of BPDs in epitaxial 4H-SiC layers [28]. Along with the experimental analysis on the generation and evolution of BPDs, as well as the efforts to reduce the density of BPDs, theoretical researches have also been conducted to understand the geometric properties, electronic properties, and kinetics of BPDs (including PDs consisting them) in 4H-SiC. It has been found that the 90° Si-core and 90° C-core PDs are stabilized at the double-period reconstruction [29]. The activation energy for the glide of the Si-core PD is lower than that of the C-core PD, indicating the expansion or shrinkage of BPDs is motivated by the glide of the Si-core PD, which gives rise to the BPD-related reliability issue in 4H-SiC-based bipolar devices [30-32]. However, the effect of BPDs on the electronic properties of 4H-SiC, and thus the electrical performance of 4H-SiC-based power devices is still ambiguous. Meanwhile, the N concentrations in N-doped 4H-SiC substrates and buffer layers are in the range of $10^{17}$-$10^{19}$ cm$^{-3}$ [28,33,34]. It is reasonable to expect that N atoms could interact with BPDs and change the properties of BPDs in 4H-SiC, while this issue has been rarely investigated. The incomplete understanding on the properties of BPDs, as well as the interaction between N and BPD in 4H-SiC impedes the understanding on the electron-transport mechanism in 4H-SiC-based power devices, and hinders further optimization on the properties of 4H-SiC.

In this work, the interaction between N atoms and BPDs, as well as its effect on the electrical properties 4H-SiC are systematically investigated combining experimental researches and first-principles calculations. After revealing BPDs via the molten-alkali etching, we demonstrate that BPDs create acceptor-like states in undoped 4H-SiC while they act as donors in N-doped 4H-SiC by the Kelvin probe force microscopy (KPFM) analysis. First principles calculations indicate that the formation energy of the BPD with the single-period (SP) reconstructed PDs is lower than that of the BPD with the double-period (DP) reconstructed PDs. BPDs create occupied defect states above the valance band maximum (VBM) and unoccupied defect states under the conduction-band minimum (CBM) of undoped 4H-SiC. Defect-formation-energy calculations demonstrate that N atoms can spontaneously decorate BPDs in N-doped 4H-SiC. The N-BPD complex is stable against decomposition because of its positive binding energy. The accumulation of N dopants at the core of BPDs results in the accumulation of donor-like states at the core of BPDs in N-doped 4H-SiC.

## II. EXPERIMENTAL METHODS

Undoped and N-doped 4H-SiC boules are grown by the physical vapor transport (PVT) approach, with the growth temperature ranging from 2000 to 2300 °C and the growth pressure ranging from 1 to 10 mbar. N doping is realized by adopting high-purity nitrogen gas ($N_2$) as the doping source, which is not added in the undoped process. The on-axis slicing and 4.0° off-axis slicing are common processing requirements of commercial undoped and N-doped 4H-SiC substrate wafers, which facilitates the subsequent heteroepitaxy (gallium nitrides on undoped 4H-SiC) and homoepitaxy (4H-SiC on N-doped 4H-SiC), respectively. In order to guarantee the universality of the current research, the slicing angles for undoped and N-doped 4H-SiC boules are the same as the common commercial ones. Lapping, Si-face CMP and cleaning are then subsequently carried out to fabricate wafers. By eddy-current resistivity measurements, the electrical resistivity of the undoped and N-doped 4H-SiC substrate wafers are found to be $9.8 \times 10^{10}$ Ω·cm and 0.02 Ω·cm, respectively (Semimap Corema-WT for high-resistivity 4H-SiC and Semilab LEI-1510 for low-resistivity 4H-SiC). The rocking-curve measurements are performed by high-resolution x-ray diffractometer (XPert3 MRD, Malvern Panalytical). Measuring 13 points taken in a cross on the wafer, the average peak positions of the undoped and N-doped 4H-SiC wafers are found to be 17.77° and 21.76°, respectively. The average FWHM of the undoped and N-doped 4H-SiC wafers are 20.13 arcsecs and 25.68 arcsecs, respectively. The concentrations of N in 4H-SiC samples are measured by the secondary ion mass spectroscope (SIMS) (IMS 4f, Cameca). The concentrations of N in undoped and N-doped 4H-SiC wafers are in the order of magnitudes of $10^{16}$ cm$^{-3}$, and $10^{18}$ cm$^{-3}$, respectively.

Molten-KOH etching of the 4H-SiC wafers are carried out at 550 °C in a Ni crucible,

128

with the 4H-SiC samples placed in a Ni-wire mesh. The etching durations for undoped and N-doped 4H-SiC samples are 20 and 30 min, respectively. The surface morphologies of etched 4H-SiC samples are characterized by optical microscope (OM, BX53M, Olympus). The surface morphologies and depth profiles of etch pits for BPDs are characterized by the atomic force microscope (AFM), (Dimension Edge, Bruker). The surface potentials of both the etch pits of BPDs and perfect regions are measured by the Peak Force KPFM equipped in the AFM with a Pt-Ir-coated Si cantilever.

## III. Results and discussion

### A. BPDs in undoped 4H-SiC and N-doped 4H-SiC

Figures 1 (a) and 1(e) display the OM images of the etched pits in undoped and N-doped 4H-SiC, respectively. Similar to previous researches, the etch pits of BPDs are characterized by sea-shell shape. The threading screwing dislocations (TSDs) and TEDs have hexagonal etch pits, with the average etch-pit size of TSDs being larger than 181 that of TEDs [6,35,36]. Figure 1(b) presents the surface profile of a typical sea-shell-shaped etch pit in the undoped 4H-SiC etched by molten KOH. The distances along the major axis [A-B in Fig. 1(b)] and minor axis [C-D in Fig. 1(b)] are similar. The bottom for the etch pit of the BPD in undoped 4H-SiC is overall flat with a deeper pit at the edge. The depth for the etch pit of the BPD is 1.1 μm, and the width spreads in the range of 20.0-22.2 μm [Fig. 1(b)]. Although undoped 4H-SiC wafers are on190 axis sliced, BPDs can be revealed when the molten-KOH etching stops at the interface where the BPD-TED conversion occurs. As illustrated in Fig. 1(d), the molten-KOH 193 etching preferentially attacks the strained surface atoms surrounding the TED by removing the stressed atoms surrounding the dislocation line of the TED. As the duration of the molten-KOH etching increases, the isotropic etching removes the 4H-SiC layer containing the TED, and reaches the interface where the BPD-TED conversion occurs. The lateral etch along the BPD is then activated by removing the strained atoms surrounding the two PDs that constitute the BPD. Therefore, the bottom of the etch pit is flat [Fig. 1(b)]. Meanwhile, we find that the flat etch pit terminates at a deeper sharp pit. This indicates that the stress field is the highest at the edge of the BPD, which gives rise to the excess preferential etching along the edge of the BPD in undoped 4H-SiC.

For N-doped 4H-SiC, sea-shell-shaped etch pits are formed after molten-KOH etching [Fig. 1(f)]. By plotting the depth profiles along different axis of the etch pit, we find the cross-section morphologies along the major axis and minor axis of the etch pit are triangle shaped (A-B) and arc shaped (C-D), respectively [Fig. 1(g)]. As illustrated in Fig. 1(h), the 4.0° off-axis slicing of N-doped 4H-SiC gives rise to the incline of the [0001] plane, and thus termination of PDs at the surface of the wafer, which facilitates the preferential etching by removing strained atoms along the dislocation lines of the PDs.

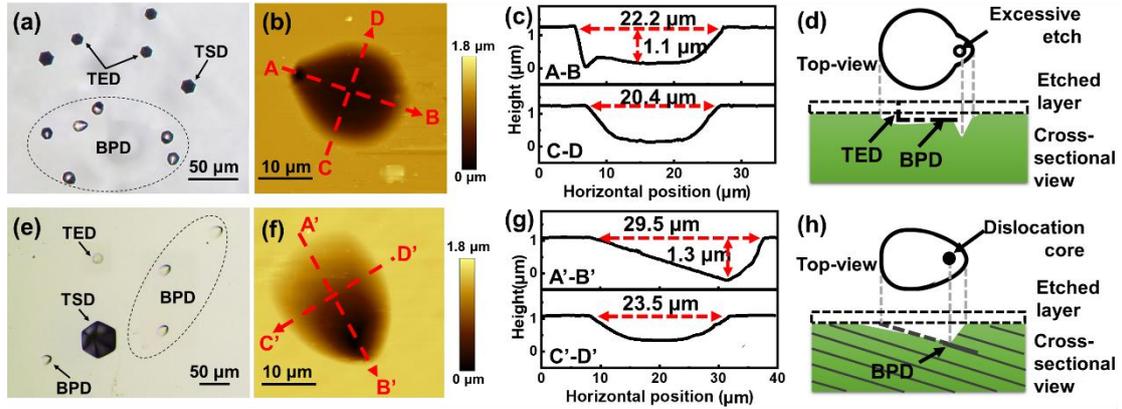

FIG. 1. Surface morphologies and depth profiles of representative etch pits of the BPDs in (a)-(c) undoped 4H-SiC, and (e)-(g) N-doped 4H-SiC. Schematic diagrams illustrating the etching mechanisms of BPDs in undoped and N-doped 4H-SiC are shown in (d) and (h), respectively.

**B. Electronic properties of BPDs**

The electronic properties of BPDs are analyzed by detecting the contact potential difference ($V_{CPD}$) between the tip and the etched 4H-SiC by the KPFM approach. The surface potential mappings of the representative BPDs in undoped and N-doped 4H-SiC are shown in Figs. 2(a) and 2(b), respectively. KPFM is capable of imaging the local work function of semiconductors by [37]

$$V_{CPD} = (\Phi_{tip} - \Phi_{sample})/-q \quad (1)$$

where $\Phi_{sample}$ and $\Phi_{tip}$ are work functions of the Pt-Ir coated tip and the sample,

respectively, and $q$ is the electronic charge. The local work function ($\Phi$) is related to the local Fermi energy ($E_F$) by

$$\Phi = \chi + (E_C - E_F) \qquad (2)$$

where $\chi$ is the electron affinity of 4H-SiC (3.80 eV) [2], $E_C$ and $E_F$ are the energies of the CBM and Fermi energy of 4H-SiC, respectively. The absolute value of the work function of the tip is calibrated by measuring the $V_{CPD}$ of a standard Au film (approximately 60 nm thickness) deposited on a Si substrate. The measured surface potential of the tip is higher than that of Au by 200 mV. With Eq. (1) and the standard work function of Au being 5.2 eV, we obtain the work function of the tip being 5.4 eV.

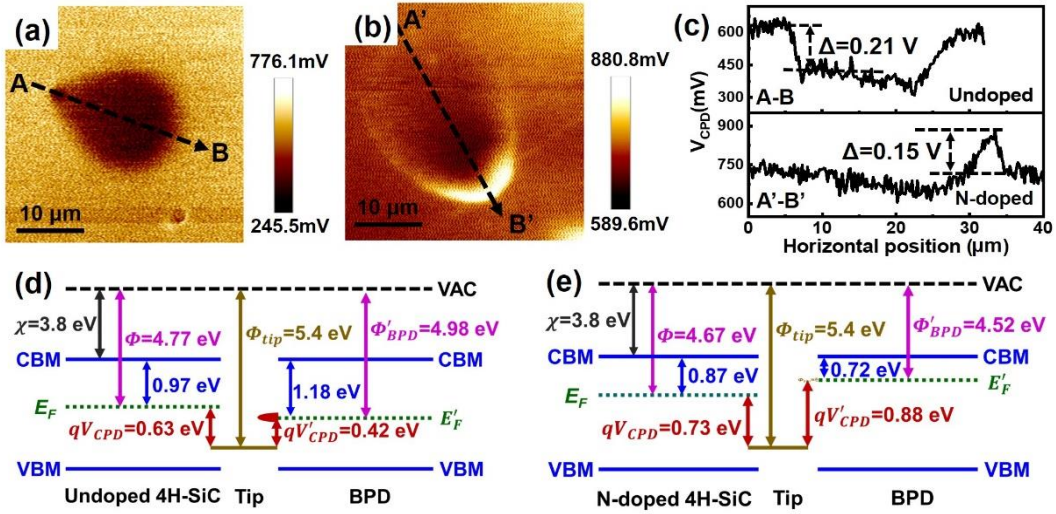

FIG. 2. Surface potential of a representative BPD in (a) undoped and (b) N-doped 4H-SiC. (c) shows the potential variations along the break lines in (a) and (b). (d) and (e) depict the band diagram of undoped and N-doped 4H-SiC containing the BPD.

As shown in Figs. 2(a) and 2(c), the local potential of a BPD in undoped 4H-SiC are lower than that of the perfect region by 0.21 V. With Eqs. (1) and (2), as well as the assistance of band-diagram analysis, we find that the local Fermi energy of the BPD in undoped 4H-SiC is 0.21 eV lower than that of the perfect region [Fig. 2(d)]. This indicates that BPDs create acceptor-like defect states in undoped 4H-SiC, the charge transfers from the perfect region to the defect states of BPDs, and thus the electron accumulation at BPDs lower the local Fermi energy of the BPD in undoped 4H-SiC.

In N-doped 4H-SiC, the local potential of the peripheral region surrounding the core of the BPD is remarkably higher than that of other regions [Figs. 2(b) and 2(c)]. The local potential for the peripheral region surrounding the core of the BPD is higher than that of other regions by 0.15 eV in N-doped 4H-SiC. With the assistance of band diagram analysis, we find that N doping makes the local Fermi level of BPDs shift to higher positions in the band gap of 4H-SiC [Fig. 2(e)].

**C. First-principles verifications**

In order to understand the interaction between N and BPDs, and their effect on the electronic properties in 4H-SiC, first-principles calculations are then carried out within the density-functional-theory (DFT) framework implemented in the Vienna ab initio simulation package (VASP) [38,39]. The core electrons are treated within the projector augmented wave (PAW) pseudopotentials, with the plane-wave cutoff for the wave-function expansion of 450 eV. BPDs are modeled in a 768-atom orthorhombic supercell, with the $x$, $y$, and $z$ axis oriented to the [11$\bar{2}$0], [1$\bar{1}$00] and [0001] directions of the hexagonal lattice of 4H-SiC, respectively. The volume of supercell is 43 Å×9 Å×20 Å. The structures of BPDs are relaxed using the Perdew-Burke-Ernzerhof (PBE) exchange-correlation functional [40]. During the structural optimizations, atomic positions are fully relaxed until the total energy per cell and the force on each atom converge to less than $1 \times 10^{-5}$ eV and 0.01 eV/Å, respectively. Since the PBE functional significantly underestimates the band-gap energies of semiconductors, the screened hybrid functional of Heyd, Scuseria, and Ernzerhof (HSE) is then employed to calculate the electronic properties of BPDs in 4H-SiC [41]. With the fraction of the screened Fock exchange of 0.26, the calculated band-gap energy of 4H-SiC is 3.19 eV, which agrees well with experimental results [2]. The Brillouin-zone integration is sampled with the Γ-centered $1 \times 2 \times 1$ Monkhorst-Pack $k$-point mesh during structural relaxations and electronic calculations [42].

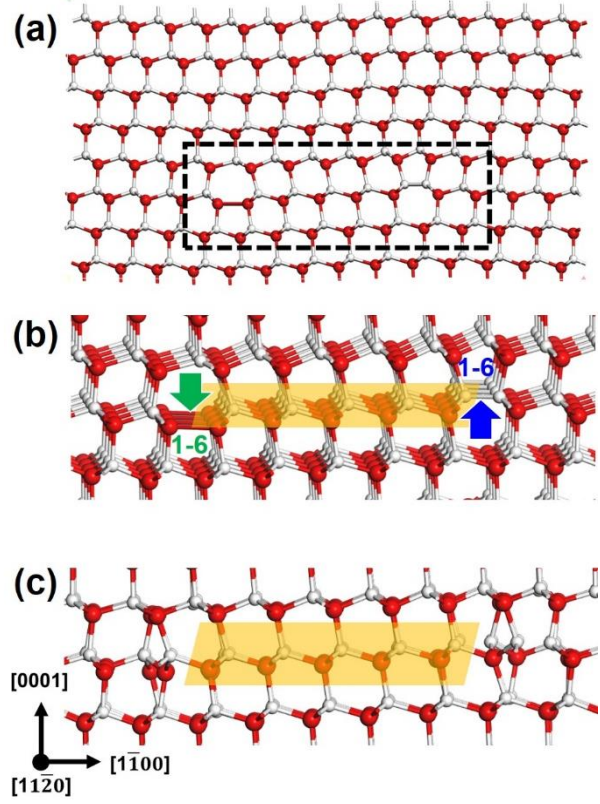

FIG. 3. (a) Relaxed 4H-SiC supercell containing a BPD. The BPDs with the SP- and DP-reconstructed PD cores are shown in (b) and (c), respectively. The yellow region, blue arrow and green arrow denote the SF, C-core PD and Si-core PD, respectively. The dislocation-core atoms are denoted by blue and green Arabic numerals, respectively. The red and white balls represent Si and C atoms, respectively.

Figure 3(a) displays the relaxed supercell containing a typical BPD in 4H-SiC, which is characterized by the formation of both Si-core PD and C-core PD separated by an SSF between them. According to the atomic reconstruction of the PD cores, the reconstruction of PDs is divided to the SP reconstruction and DP reconstruction. The SP-reconstructed PD core is constructed by directly connecting the bulk region with the stacking fault region [Fig. 3(b)], and the DP-reconstructed PD core is constructed from the SP reconstruction by introducing double kinks at alternating sites along the PD core [Fig. 3(c)] [43]. We firstly calculate the formation energy of BPD [$\Delta H_f(BPD)$] by

$$\Delta H_f(BPD) = E_t(BPD) - E_t(host) + \sum n_i(\mu_i + E_i) \qquad (3)$$

where $E_t(BPD)$ and $E_t(host)$ are total energies of the 4H-SiC host with and without the

BPD, respectively. $n_i$ is the number of constituent $i$ transferred from the supercell to the reservoir during the formation of the BPD, $\mu_i$ is the chemical potential of constituent $i$ referenced to its elemental phase with energy $E_i$. The sum of $\mu_{Si}$ and $\mu_C$ is limited by the total energy of bulk 4H-SiC to maintain the host in equilibrium. The individual values of $\mu_{Si}$, $\mu_C$ and $\mu_N$ are limited by the total energy per atom of bulk Si, bulk C and $N_2$ to avoid the elemental precipitation. The chemical potentials of N and Si are also limited by the formation energies of $Si_3N_4$ to avoid the formation of the secondary phase of $Si_3N_4$. Consistent with the growth condition of 4H-SiC, the Si-rich limit with $\mu_{Si}$ =0 eV is adopted.

Firstly, we compare the relative stabilities of BPDs containing PDs with the SP and DP reconstructions. For the PD with a Si core or a C core, it is found that the formation energy of the DP-reconstructed PD is lower than that of the SP-reconstructed PD, because of the elimination of dangling bonds at the PD cores [29]. However, for the BPD consisting of two PDs separated by a SSF, we find that the formation energy of the BPD with SP-reconstructed PDs is lower than that of the BPD with DP-reconstructed PDs by 0.39 eV. Considering the dislocation length of 12.3 Å, the formation energy of the BPD with SP-reconstructed PDs is lower by 0.02 eV/Å than that of the BPD with DP reconstructed PDs. We note that in the current model, the area of the SSF is 12.3 Å × 10.7 Å, which would affect the formation energy of the whole BPD. A further study on the effect of the SSF width and the length of PDs on the formation energy of the whole BPD will be done in the future. The above calculations indicate that when the formation of the SSF is taken into account, the BPD with SP-reconstructed PDs is more energetically stable because of the less lattice distortion. Therefore, we use the configuration of the BPD with SP-reconstructed PDs separated by a SSF to investigate the interaction between N and BPDs in 4H-SiC.

Since the close-packed {0001} planes are stacked in sequence of *ABCB* in 4H-SiC, the (0001)-located SSF can be created at various stacking planes. It turns out the formation-energy difference between BPDs created at different stacking planes is smaller than 0.03 eV, which indicates that the BPD can be formed equally at different stacking planes of 4H-SiC.

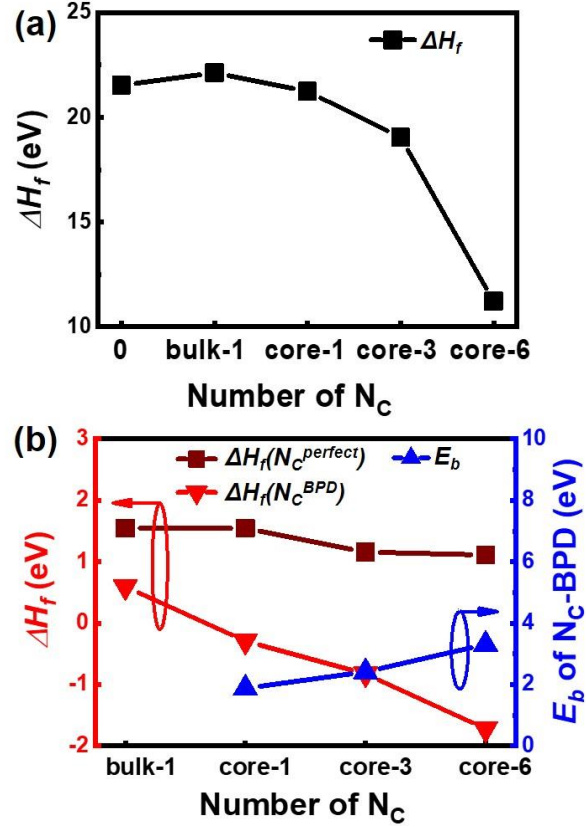

FIG. 4. (a) Calculated formation energies ($\Delta H_f$) of the pure BPD, as well as BPDs decorated by $N_C$ located at the perfect region and the core of BPD in 4H-SiC. (b) Calculated formation energies ($\Delta H_f$) of $N_C$ in a 4H-SiC supercell with [$\Delta H_f\left(N_C^{BPD}\right)$] and without the BPD[$\Delta H_f\left(N_C^{perfect}\right)$], as well as binding energies ($E_b$) of $N_C$-BPD as functions of the number of $N_C$ at the BPD core, the formation energies of $N_C$ in perfect 4H-SiC are enclosed for comparison. (c) Schematic diagram showing the interaction between $N_C$ and BPD. Electrons and holes are denoted by closed and open balls, respectively.

During N doping, the N substitution of C ($N_C$) can happen either at the core of the BPD or inside the bulk region of 4H-SiC. The calculated formation energies of the BPDs with and without NC decoration are shown in Fig. 4(a). The formation energy of the BPD in 4H-SiC is 21.5 eV. Creating an additional $N_C$ in the perfect region away from the BPD increases the formation energy of BPD in 4H-SiC. However, when $N_C$ is created at the core of the BPD, the formation energy of the $nN_C$-BPD ($n$=1, 3, 6) continuously decreases. This indicates that N atoms tend to substitute C atoms at the core of BPD during N doping of 4H-SiC. It should be noted that we also compare the formation energies of $N_C$ and interstitial N ($N_i$) at the core of the BPD. The calculated formation energy of $N_i$ at the core of the BPD is 23.14 eV, which is higher than those of $N_C$ located at the core of the BPD (21.24 eV) and in the bulk region without defect (22.13 eV). This verifies that $N_C$ is the dominant defect configuration of N in 4H-SiC containing the BPD. In order to verify the preference of $N_C$ at the core of BPDs, we then calculate the formation energy of $N_C$ at the supercell of 4H-SiC containing a BPD. The positive value of NC located in the perfect region away from the BPD indicates that it costs energy to substitute C atoms at the perfect region away from the BPD of 4H-SiC. However, when N substitutes C at the core of the BPD, the value of formation energy of NC becomes negative. When the number of N atoms decorating the BPD core increases from 1 to the maximum of 6 in this supercell, the formation energy of $N_C$ continuously decreases [Fig. 4(b)]. Therefore, it is reasonable to deduce that N atoms tend to substitute all C atoms at the core of BPDs. We also compare the formation energy of $N_C$ in a perfect 4H-SiC supercell without any defects, and that in a 4H-SiC supercell containing a BPD. As shown in Fig. 4(b), the formation energies of $N_C$ in a 4H-SiC supercell containing a BPD [$\Delta H_f(N_C^{BPD})$] are always lower than those of $N_C$ in aperfect 4H-SiC supercell [$\Delta H_f(N_C^{perfect})$]. This verifies that N atoms prefer to decorate the cores of BPDs during the N doping of 4H-SiC. Only when the decoration becomes saturated, N atoms begin to substitute C at the perfect region of 4H-SiC.

In order to verify the decoration of N at BPDs, we then calculate the binding energy ($E_b$) of the $nN_C$-BPD ($n$=1-6) complex by:

$$E_b(nN_C - BPD) = \left[n\Delta H_f(N_C) + \Delta H_f(BPD) - \Delta H_f(nN_C - BPD)\right]/n \qquad (4)$$

As shown in Fig. 4(b), the positive values of $E_b$ indicate that once N atoms decorate BPDs, the $nN_C$-BPD complex is strong against decomposition. Furthermore, we find 399 that the binding energy of the $nN_C$-BPD (n = 1-6) complex increases when the value of n increases. This indicates that even the core of the BPD is fully decorated by N, the $N_C$-BPD complex is stable against decomposition.

The physical origin for the accumulation of $N_C$ at the core of BPDs can be understood by electronic interaction between the defect states of $N_C$ and the BPD. Eigenvalue analysis indicates that the BPD creates occupied states above the VBM and unoccupied states below the CBM of 4H-SiC. Because of the reduced symmetry of 4H-SiC containing a BPD, the defect states of the BPD are *a* states [Fig. 4 (c)]. We note that the defect states of BPDs are continuum states rather than a single state. The defect state of NC caused by the N donor is shallow and below the CBM of 4H-SiC, which is occupied by one electron. Because the symmetries of defects states of NC and BPD are the same, the coupling between them pushes the defect state of NC to higher positions and those of the BPD to lower positions. The single electron from the original shallow donor level of $N_C$ then transfers to the unoccupied states of the BPD. The energy gain of the electron-transfer process gives rise to the accumulation of NC at the core of BPDs in 4H-SiC.

Next, we calculate the density of states (DOSs) of BPDs in 4H-SiC. As shown in Fig. 5(a), the BPD creates two defect levels in the band gap of 4H-SiC. Single-particle 423 analysis indicates that the defect level above the VBM of 4H-SiC is fully occupied by electrons, while the defect level below the CBM is empty. This agrees well with the above physical picture about the defect states of the BPD in 4H-SiC. In undoped 4H-SiC, the semi-insulating character is caused by the vacancy clusters, which create the (+/0) transition level locating at 1.8 eV below the CBM [44,45]. Furthermore, the concentration of native impurities, such as V, Cr, Fe, Co, Ni, and S, are comparable to that of vacancy clusters [2]. The native impurities create various defect states in the band gap of 4H-SiC [46]. As illustrated in Fig. 5(b), the empty defect level of BPD under the CBM of 4H-SiC is capable of accepting electrons excited from the midgap states of intrinsic defects and native impurities. The electron transfer from intrinsic defects or native impurities to the empty defect level of BPD gives rise to the acceptorlike behavior of BPDs in undoped 4H-SiC [Fig. 5(b)].

For N-doped 4H-SiC, the concentration of N is usually in the range of $10^{18}$-$10^{19}$ cm$^{-3}$ [33,47], the density of BPDs is usually ranging from $10^2$ cm$^{-2}$ to $10^3$ cm$^{-2}$ [23, 48-50]. The density of BPDs is much lower than that of N dopants, meanwhile, formation-energy and binding-energy calculations indicate that N dopants tend to spontaneously accumulate at the core of BPDs. Therefore, the configuration of $N_C$-BPD in N-doped 4H-SiC is energetically favored, which is characterized by the six N substituting the C atoms at the C-core PD consisting the BPD ($6N_C$-BPD) in this supercell. As shown in Fig. 5(c), the defect level above the VBM of 4H-SiC basically keeps the same upon fully NC decoration. Meanwhile, two separate defect levels are created under the CBM of 4H-SiC. It should be noted that the calculated DOS is the electronic properties of the NC-decorated BPD in 4H-SiC. Therefore, the single $N_C$-BPD might be more n-type doped than the single BPD. In KPFM measurements, the local Fermi energy of the BPD is compared with the perfect region of 4H-SiC. In N-doped 4H-SiC, because N dopants prefer to accumulate at the core of BPDs, the density of N dopants around BPDs is higher than that at the perfect region of 4H-SiC. The N dopants rise the local Fermi energy up, and give rise to the donor-like behavior of the $N_C$-decorated BPDs in N-doped 4H-SiC. The excess electron of $N_C$ at the core of the BPD would either transfer to the CBM, or to the perfect region where high density of vacancy clusters also exists [Fig. 5(d)]. We note that the N decoration of BPDs can happen during the N-doping process of 4H-SiC. That is, during the doping process of N, once the N atoms encounter the thermal-stress-induced BPD, they could spontaneously decorate the BPDs. Furthermore, the growth temperatures of 4H-SiC substrates and epitaxial layers are as high as 2000–2300 °C and 1500–1700 °C, respectively. The high temperature also provides sufficient energy for the diffusion of N atoms.

At last, we discuss the effect of N decoration of BPDs in 4H-SiC on the electronic properties of 4H-SiC-based power devices. For reversely biased 4H-SiC devices, the dominant emission mechanism is the Poole-Frenkel (PF) emission, which is featured by the transfer of electrons through the continuum defect states under CBM [51,52]. In N-doped 4H-SiC, the N decoration results in two unoccupied defect levels under the CBM [Fig. 5(c)]. These two defect levels may serve as electron transfer channels, which would increase the leakage current of 4H-SiC-based power devices under the reverse-bias condition. Therefore, BPDs not only give rise to the degradation of bipolar devices, but also may be involved in the reverse leakage current of N-doped 4H-SiC devices.

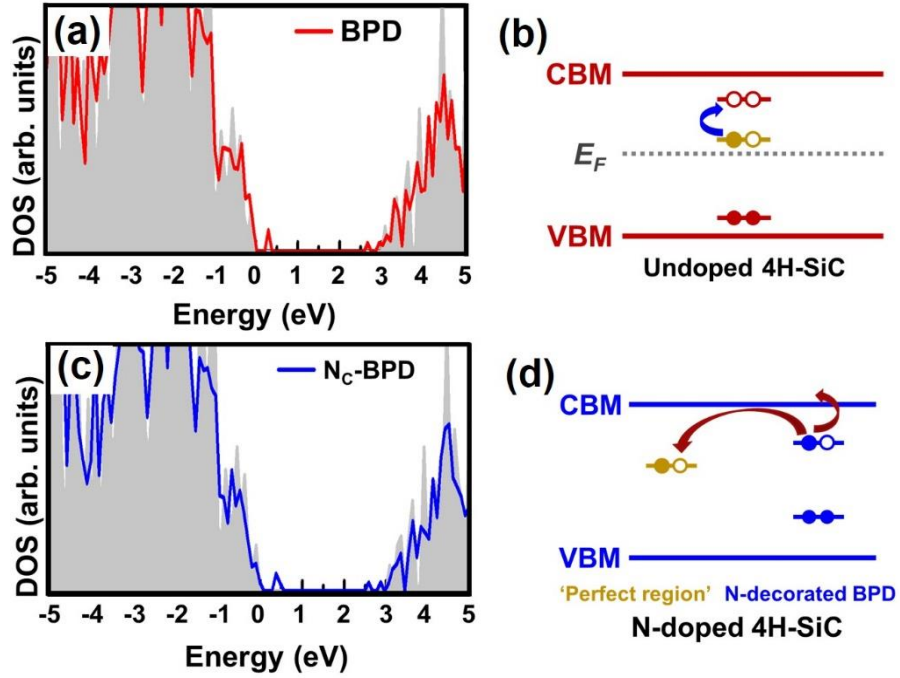

FIG. 5. Calculated DOSs of (a) the pure BPD, and (c) $N_C$-decorated BPD in 4H-SiC. The DOSs of the bulk 4H-SiC is indicated by the gray regions. Band diagrams showing the acceptor-like behavior of BPD in undoped 4H-SiC and donor-like behavior of N-decorated BPD in N-doped 4H-SiC are shown in (b) and (d), respectively.

## IV. CONCLUSIONS

In conclusion, we systematically explore the electronic properties of BPDs in undoped and N-doped 4H-SiC. By investigating the surface and cross-section morphologies for the etch pits of BPDs after molten-KOH etching, we establish the etching mechanism of BPDs in on-axis sliced undoped 4H-SiC and off-axis sliced N-doped 4H-SiC. KPFM measurements indicate that BPDs create acceptor-like states in undoped 4H-SiC while they act as donors in N-doped 4H-SiC. First-principles calculations verify that BPDs create defect states both above the VBM and below the CBM of 4H-SiC. The electron transfer from the defect states of intrinsic defects and native impurities to the unoccupied defect states results in the acceptor-like behavior of BPDs in undoped 4H-SiC. In N-doped 4H-SiC, we find that N atoms prefer to decorate the core of BPDs. The accumulation of N dopants gives rise to the donor-like behavior of BPDs in N-doped 4H-SiC. Our work not only reveals the role of BPDs on the electronic properties of 4H-SiC, but also paves the way for understanding the electron transport mechanism in 4H-SiC-based power devices.

**ACKNOWLEDGEMENTS**

This work is supported by "Pioneer" and "Leading Goose" R&D Program of Zhejiang Province (Grant No. 2022C01021), Natural Science Foundation of China (Grants No. 91964107, No. U20A20209), Natural Science Foundation of China for Innovative Research Groups. (Grant No. 61721005). National Supercomputer Center in Tianjin is acknowledged for computational support.